\pdfoutput=1

\documentclass[10pt]{iopart}
\usepackage{iopams}  
\usepackage{hyperref}
\usepackage{color}

\usepackage{graphicx}
\usepackage[a4paper]{anysize}%

\definecolor{darkred}{rgb}{0.5,0.0,0.0}
\definecolor{darkgreen}{rgb}{0.0,0.5,0.0}
\definecolor{darkblue}{rgb}{0.0,0.0,0.5}

\hypersetup{colorlinks=true,
citecolor=darkred,
linkcolor=darkgreen,
urlcolor=darkblue}

\marginsize{3cm}{2.5cm}{2.0cm}{4.0cm}

\newcommand{\abbr}[1]{#1}
\newcommand{\lloc}{\ensuremath{\ell_{\mathrm{loc}}}}
\newcommand{\lel}{\ensuremath{\ell_{\mathrm{el}}}}
\newcommand{\Nch}{\ensuremath{N_{\mathrm{ch}}}}
\newcommand{\kF}{\ensuremath{k_{\mathrm{F}}}}

\newcommand{\tmfloatsmall}[2]{
\begin{figure}
#1
\caption{#2}
\end{figure}}

\newcommand{\tmfloatfirst}[2]{
\begin{figure}
#1
\caption{#2}
\end{figure}}

\begin{document}

\title{Diffusion and localization in carbon nanotubes and graphene
nanoribbons}

\author{Norbert Nemec}
\address{Institute for Theoretical Physics,
 University of Regensburg,
 D-93040 Regensburg, Germany}
\address{TCM Group,
 Cavendish Laboratory,
 University of Cambridge,
 Cambridge CB3 0HE, UK}

\author{Klaus Richter}
\address{Institute for Theoretical Physics,
 University of Regensburg,
 D-93040 Regensburg, Germany}

\author{Gianaurelio Cuniberti}
\address{Institute for Theoretical Physics,
 University of Regensburg,
 D-93040 Regensburg, Germany}
\address{Max Bergmann Center for Biomaterials,
 Dresden University of Technology,
 D-01062 Dresden, Germany}

\ead{g.cuniberti@tu-dresden.de}

\begin{abstract}
  We study transport length scales in carbon nanotubes and graphene ribbons
  under the influence of Anderson disorder. We present generalized analytical
  expressions for the density of states, the elastic mean free path and the
  localization length in arbitrarily structured quantum wires. These allow us
  to analyze the electrical response over the full energy range, including the
  regions around van Hove singularies, traditionally difficult to access by
  alternative approaches. Comparing with the results of numerical simulations,
  we demonstrate that both the diffusive and the localized regime are well
  represented by the analytical approximations over a wide range of the energy
  spectrum. The approach works well for both metallic and semiconducting
  nanotubes and nanoribbons but breaks down near the edge states of zigzag
  ribbons.
\end{abstract}



\maketitle


The exceptionally high electrical conductivity of carbon nanotubes
({\abbr{CNT}}s) can be attributed to a combination of various factors. The
quasi-one-dimensional (quasi-{\abbr{1D}}) crystalline structure allows the
preparation of samples with a very low defect
density\ \cite{fan-iacpdicn2005}. The rigidity of the $s 
p^2$-hybridized carbon lattice minimizes the effects of thermal
vibrations\ \cite{suzuura-paesicn2002}. One truely unique feature of metallic
{\abbr{CNT}}s, however, is their electronic
structure\ \cite{saito-esogtboc1992}: independent of the tube diameter, there
are exactly two massless bands of high velocity crossing at the Fermi energy,
resulting in a very low density of states ({\abbr{DOS}}) which effectively
suppresses scattering even in the presence of
disorder\ \cite{white-cnalbc1998} and allows ballistic transport over hundreds
of nanometers\ \cite{mann-btimnwrpoc2003,purewal-soraemfposcn2007}.

Planar, finite-width graphene nanoribbons ({\abbr{GNR}}s) display many
similarties to their rolled-up counterparts. The band gap of armchair-edge
{\abbr{GNR}}s depends on the width just as it depends on the circumference
in zigzag {\abbr{CNT}}s. Zigzag-edge {\abbr{GNR}}s, on the other hand, are
always metallic just as armchair {\abbr{CNT}}s. A remarkable difference,
however, lies in the edge states generally found at zigzag edges in
graphene\ \cite{nakada-esigrnseaesd1996,fujita-plsazge1996} that are nearly
localized by their extremely low dispersion and exist beside the conducting
channels at the band center\ \cite{sasaki-smoesig2006}. Energetically, these
states lie exactly around the Fermi energy, so it would be expected to have an
important impact on the low-energy transport properties of {\abbr{GNR}}s.
The low velocity of the edge channels results in a large {\abbr{DOS}}, which
should by itself lead to strong scattering. However, the local {\abbr{DOS}}
({\abbr{LDOS}}) is strongly concentrated on the edge atoms and protecting
the conduction channel from scattering into the edge states.

The theory of quantum transport in disordered systems has been studied for
many decades\ \cite{anderson-aodicrl1958}. Typically, the underlying model for
such studies is that of free electrons or, if a discretization is desired,
electrons on a (square) lattice. Quantities such as effective electron mass,
density of states or Fermi velocity are then viewed as free parameters that
can be adjusted to fit the properties of metals\ \cite{mott-ttoic1961}.
Graphene and {\abbr{CNT}}s, however, have an electronic structure that is
very different from any effective-mass free-electron approximation. In
two-dimensional graphene, the electrons have a Dirac-like massless
dispersion\ \cite{wallace-tbtog1947,novoselov-tgomdfig2005}, which results in
equally massless bands of high velocity in metallic
{\abbr{CNT}}s\ \cite{saito-esogtboc1992}. Tuning the Fermi energy away from
the charge neutrality point ({\abbr{CNP}}) via electronic gating or chemical
doping, {\abbr{CNT}}s of all chiralities go through a sequence of additional
bands that lead to van-Hove singularities in the {\abbr{DOS}} which are
typically not covered by simple models based on a smooth or fixed
{\abbr{DOS}} and number of conductance channels.

In this paper, we will present a generalized expression for the {\abbr{DOS}}
in arbitrarily structured, Anderson-disordered quantum wires that allows to
access the whole energy range including the regions around the van Hove
singularities. Using this representation of the {\abbr{DOS}}, we then show
that a general expression, originally devised for metallic quantum wires with
a fixed number of channels still holds true near band edges if used correctly.
The various expressions will be applied to both {\abbr{CNT}}s and
{\abbr{GNR}}s, including the special case of pure edge disorder in the
latter case.

For {\abbr{1D}} systems such as {\abbr{CNT}}s, it was argued by Mott and
Twose\ \cite{mott-ttoic1961} that all quantum states should become localized
at arbitrarily low disorder, leading to an exponential suppression of the
conductance at zero-temperature. For truly {\abbr{1D}} quantum wires,
Thouless demonstrated that the localization length $\lloc$ is identical to the
mean free path $\lel$\ \cite{thouless-ldamfpiods1973}. Later, he derived a
generalized expression for metallic quantum wires of finite
diameter\ \cite{thouless-mmritw1977},
\begin{eqnarray}
  \lloc & \approx & \frac{2 A \kF^2}{3 \pi^2} \lel, 
\end{eqnarray}
where the prefactor consisting of cross section $A$ and Fermi wave vector
$\kF$ can be identified as the number of conduction channels $\Nch$ in the
modern language of quantum transport theory. A more accurate expression
derived by Beenakker from random matrix theory
reads\ \cite{beenakker-rtoqt1997}
\begin{eqnarray}
  \lloc & \approx & \left[ \beta \left( \Nch - 1 \right) / 2 + 1 \right] \lel,
  \label{l_loc-Beenakker}
\end{eqnarray}
where $\beta = 1$ for time-reversal invariant systems and $\beta = 2$
otherwise. In either case, however, $\ell_{\mathrm{el}}$ and $N_{\mathrm{ch}}$ are
viewed as fixed parameters and it is not at all clear, how these relations
should be applied to systems where these quantities vary strongly with the
energy and may not even be well defined near band edges where $N_{\mathrm{ch}}$
is discontinuous. An alternative approach based on a perturbative expression
for the Lyapunov exponents takes into account the true band structure of a
quantum wire, but it still fails to reproduce the correct behavior near band
edges\ \cite{rmer-weflloqs2004}.



We model {\abbr{CNT}}s based on an orthogonal first-nearest-neighbor
tight-binding approximation\ \cite{saito-esogtboc1992} with the hopping
parameter $\gamma_0 = 2.66 \hspace{0.2em} \mathrm{eV}$, which accurately
reproduces the Fermi velocity of armchair {\abbr{CNT}}s and gives reasonable
agreement with the correct band structure for several low energy bands.

For zigzag {\abbr{GNR}}s, this approximation is insufficient, as it results
in a completely dispersion-free edge state while precise calculations show
that this state in fact has a small but non-vanishing band
width\ \cite{miyamoto-fsoesohgr1999,sasaki-smoesig2006}. Likewise, this
approximation predicts every third armchair {\abbr{GNR}} to be metallic,
while precise calculations predict the opening a small
gap\ \cite{son-egign2006,white-hoiicnrign2007}. The true electronic structure
of a {\abbr{GNR}} would have to take into account the relaxed electronic
structure at the edges\ \cite{finkenstadt-fgtgagtafnct2007}. As it turns out,
a third-nearest neighbor parameterization\ \cite{reich-tdog2002} of an
unrelaxed {\abbr{GNR}} gives a good approximation of the bands near the
Fermi energy and will thus be used here.

\tmfloatfirst{\resizebox{.7\columnwidth}{!}{\includegraphics{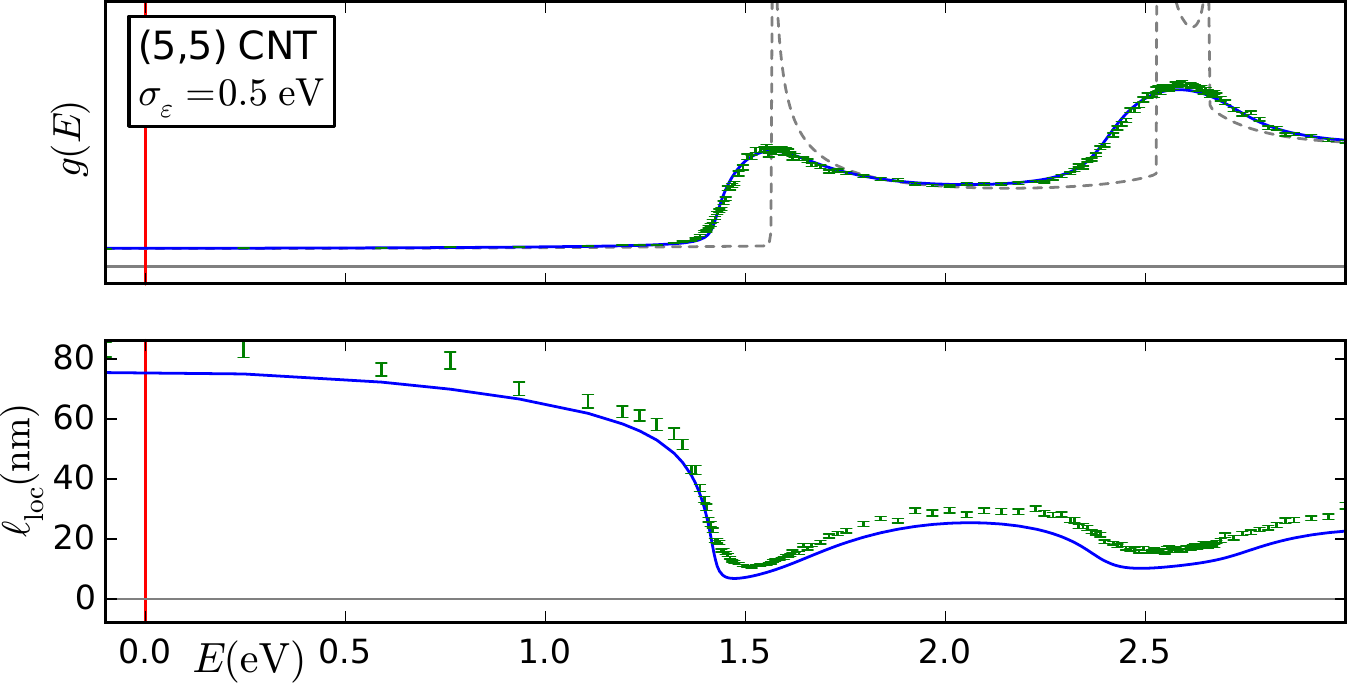}}}{\label{fig:CNT-numerically}Density
of states $g (E)$ and localization length $\ell_{\mathrm{loc}}$ of an armchair
(5,5)~{\abbr{CNT}} under the influence of Anderson disorder. Dashed line: $g
\left( E \right)$ of the clean system displaying the van Hove singularities.
Solid lines: $g \left( E \right)$ from Eq.~(\ref{g(E)}) and
$\ell_{\mathrm{loc}}$ obtained from it via Eqs.~(\ref{l_loc-Beenakker}) and
(\ref{l_el}). Data with errorbars: values obtained numerically by averaging
over 100 samples of length $1000$ to $5000 \ell_{\mathrm{uc}}$.}


Anderson disorder is defined as uncorrelated potential
fluctations\ \cite{anderson-aodicrl1958}:
\begin{eqnarray}
  \left\langle \varepsilon_i \varepsilon_j \right\rangle - \left\langle
  \varepsilon_i \right\rangle \left\langle \varepsilon_j \right\rangle & = &
  \delta_{i  j} \sigma_{\varepsilon_i}^2, 
  \label{eq:anderson-distribution}
\end{eqnarray}
usually in terms of a uniform random distribution $\varepsilon_i -
\varepsilon_0 \in \left[ - W / 2, W / 2 \right]$, which has a standard
deviation $\sigma_{\varepsilon} = W / \sqrt{12}$. For perturbatively weak
disorder, however, the exact shape of the distribution is irrelevant.
Numerically, we use a Gaussian distribution for practical reasons.
Experimentally measured values of $\ell_{\mathrm{el}}$ are of the order of $10
\hspace{0.2em} \mu \mathrm{m}$ in CNT of $1.5 \hspace{0.2em} \mathrm{nm}$
diameter\ \cite{purewal-soraemfposcn2007}, which translates to an equivalent
model disorder of strength $\sigma_{\varepsilon} \sim 0.05 \hspace{0.2em}
\mathrm{eV}$. The values used in Figs.~\ref{fig:CNT-numerically} and
\ref{fig:GNR-numerically} are chosen significantly larger to enhance the
visibility of the qualitative effects.


The diffusive transmission through a quantum wire with a disordered section of
length $L$, embedded in an otherwise disorder-free, infinite quantum wire, is
defined as
\begin{eqnarray}
  T_{\mathrm{diff}} & = & N_{\mathrm{ch}} \left( 1 + \frac{L}{\ell_{\mathrm{el}}}
  \right)^{- 1},  \label{eq:T-diff}
\end{eqnarray}
with the elastic mean free path $\ell_{\mathrm{el}}$. For armchair
{\abbr{CNT}}s at the Fermi energy, this length $\ell_{\mathrm{el}}$ was first
derived by White and Torodov\ \cite{white-cnalbc1998}. We have previously
given a generalized derivation that is valid for arbitrary
energies\ \cite{nemec-qticn2007}, which can be further generalized to also
cover the case of inequivalent atoms in the unit cell:

Within a statistical ensemble $P \left( \mathcal{W} \right)$ of the
perturbation Hamiltonian $\mathcal{W}$, the scattering rate between two bands
$s$ and $d$ at a fixed energy $E$ is given by the Fermi golden rule
\begin{eqnarray}
  \tau_{E, s \rightarrow d}^{- 1} & = & \frac{2 \pi}{\hbar} g \left( E, d
  \right) \int \mathrm{d} \mathcal{W}P \left( \mathcal{W} \right) \left|
  \left\langle E, d | \mathcal{W} | E, s \right\rangle \right|^2, 
\end{eqnarray}
which depends on the partial {\abbr{DOS}} of the destination band $d$. In
the atomic basis $\left| i \right\rangle$, the disorder Hamiltonian is
diagonal and uncorrelated [Eq.~(\ref{eq:anderson-distribution})], allowing for
the direct evaluation of the integral as
\begin{eqnarray}
  \tau_{E, s \rightarrow d}^{- 1} & = & \frac{2 \pi}{\hbar} g \left( E, d
  \right)  \left| \left\langle E, d | i \right\rangle \right|^2
  \sigma^2_{\varepsilon_i}  \left| \left\langle i | E, s \right\rangle
  \right|^2 . 
\end{eqnarray}
The elastic mean free path $\ell_{\mathrm{el}}^s$ within a band $s$ is related
to the scattering rate $\tau_{E, s}^{- 1}$ from this band via its electron
velocity which is proportional to the inverse of the {\abbr{DOS}} as
\begin{eqnarray}
  \ell_{\mathrm{el}}^s & = & v_s / \left( \sum_d \tau_{E, s}^{- 1} \right)
  \nonumber\\
  & = & \left( hg \left( E, s \right) \sum_d \tau_{E, s \rightarrow d}^{- 1}
  \right)^{- 1} . 
\end{eqnarray}
\tmfloatsmall{\resizebox{.5\columnwidth}{!}{\includegraphics{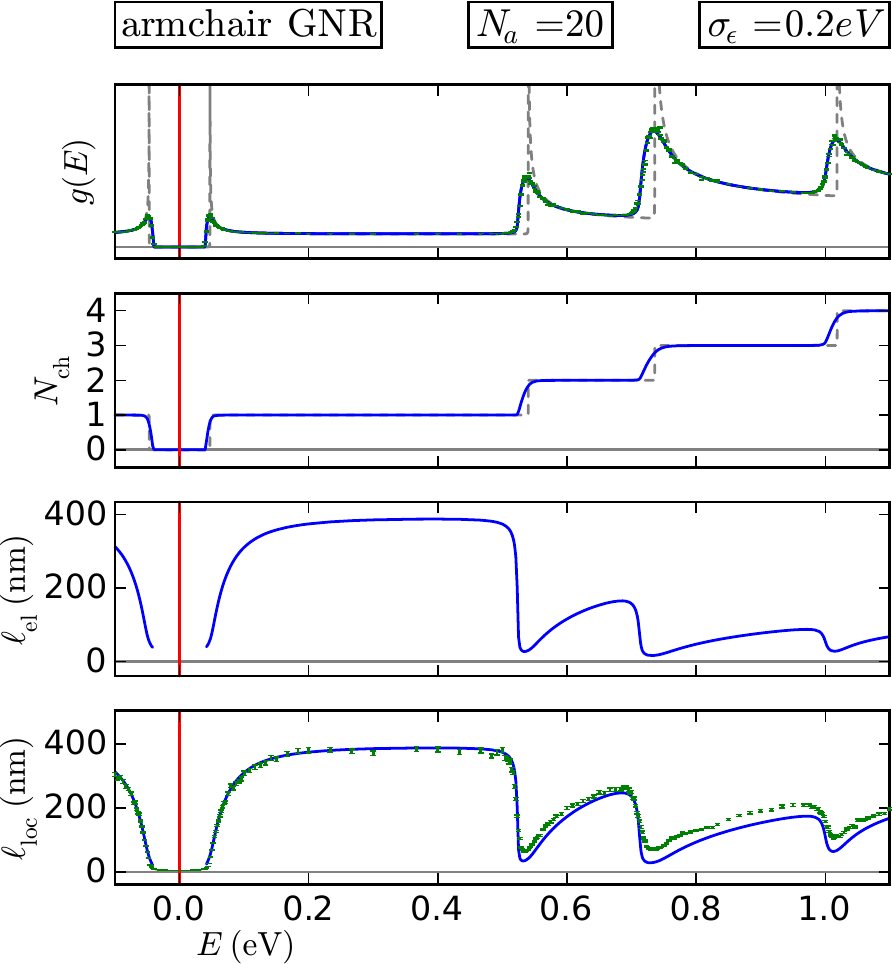}}}{\label{fig:GNR-numerically}Density
of states $g (E)$, number of channels $N_{\mathrm{ch}}$, elastic mean free path
$\ell_{\mathrm{el}}$ and localization length $\ell_{\mathrm{loc}}$ of an armchair
GNR of width $N_a = 20$ under the influence of Anderson disorder
($\sigma_{\varepsilon} = 0.2 \hspace{.2em} \mathrm{\mathrm{eV}}$). Dashed lines:
$g \left( E \right)$ and $N_{\mathrm{ch}}$ of the clean system and lengths
obtained from these via Eqs.~(\ref{l_loc-Beenakker}) and (\ref{l_el}). Solid
lines: $g \left( E \right)$ from Eq.~(\ref{g(E)}), $N_{\mathrm{ch}}$ from
Eq.~(\ref{transmission}) and lengths obtained from these. Data with errorbars:
values obtained numerically by averaging over $\sim 180$ samples of length
$2000$ to $20000 \ell_{\mathrm{uc}}$.}

Based on the definition of $\ell_{\mathrm{el}}$ via the diffusive transmission
in Eq.~(\ref{eq:T-diff}), the elastic mean free path $\ell_{\mathrm{el}}$ of a
multichannel quantum wire is found by inverse averaging
\begin{eqnarray}
  l_{\mathrm{el}}^{- 1} & = & \frac{1}{N_{\mathrm{ch}}} \sum_s \left(
  \ell_{\mathrm{el}}^s \right)^{- 1} . 
\end{eqnarray}
At this point, the sums over the bands $s$ and $d$ can be reduced, introducing
the {\abbr{LDOS}} on individual orbitals $g_i \left( E) \right.$ and
resulting in the final expression
\begin{eqnarray}
  \ell_{\mathrm{el}} & = & \ell_{\mathrm{uc}} N_{\mathrm{ch}} \left( \pi^2
  \sum_i^{\mathrm{uc}} \left( \sigma_{\varepsilon_i}^2 g_i^2 \left( E \right)
  \right) \right)^{- 1}  \label{l_el}
\end{eqnarray}
with the length of the unit cell $\ell_{\mathrm{uc}}$ and the sum running over
all orbitals $i$ within one unit cell. In this form, the expression can be
applied to arbitrary quantum wires, including {\abbr{GNR}}s, where it also
covers the special case of edge disorder by making $\sigma_{\varepsilon_i}^2$
dependent on the orbital number $i$.

Neglecting multiple scattering, the elastic mean free path $\ell_{\mathrm{el}}$
and the diffusive transmission $T_{\mathrm{diff}}$ are defined in term of the
{\abbr{LDOS}} $g_i \left( E \right)$ of the disorder-free system. Likewise,
$N_{\mathrm{ch}}$ is defined by the leads, where it follows an exact integer
step function. Near band edges, this diffusive transmission is discontinuous,
as can be confirmed numerically to arbitrary precision, computing it as the
sample average $\left\langle T \right\rangle$ of the transmission of many
disorder configurations\ \cite{nemec-qticn2007}.

The diffusion coefficient, as it can be obtained via the time evolution of a
wave packet within a long, disordered quantum wire, also allows the extraction
of the elastic mean free path\ \cite{triozon-eticnrodahs2004}. In this case,
however, it does not depend on the {\abbr{DOS}} not of the clean, but of the
disordered system. Near band edges, the {\abbr{DOS}} depends on the disorder
strength non-perturbatively, causing the van-Hove singularities to broaden and
to shift\ \cite{hgle-vhsidmqwan2002}. This effect has to be taken into account
when describing diffusion near band edges or around the edge state in
{\abbr{GNR}}s.


The density of states ({\abbr{DOS}}) of a quantum wire under the influence
of Anderson disorder can be obtained via an algorithm based on diagrammatic
perturbation theory that takes into account localization effects by including
multiple scattering\ \cite{hgle-vhsidmqwan2002}. Dropping the crossing
diagrams within the noncrossing approximation
({\abbr{NCA}})\ \cite{abrikosov-qftmisp1965}, allows to write the self
energy $\Sigma \left( E \right)$ to all orders as a recursive expression,
which can then be iterated numerically until self-consistency is reached.
Though the applicability of the {\abbr{NCA}} is not obvious, it be justified
by comparing the contribution of various terms at low
orders\ \cite{hgle-vhsidmqwan2002}.

The original formulation of this approach is restricted to the special case of
{\abbr{CNT}}s where all atoms are equivalent through symmetry and the self
energy takes the same value for all atoms. It can, however, be generalized to
arbitrarily structured quantum wires using matrix notation. The self energy
$\Sigma \left( E \right)$ caused by the disorder is a diagonal matrix obeying
the recursive relation
\begin{eqnarray}
  \left[ \Sigma \left( E \right) \right]_{i, j} & = & \delta_{i  j}
  \sigma_{\varepsilon_i}^2 \left[ \left( E + \mathrm{i} 0^+ -\mathcal{H}_0 -
  \Sigma \left( E \right) \right)^{- 1} \right]_{i  j} . 
  \label{self-energy}
\end{eqnarray}
For a periodic system, the self energy has the same periodicity as the
Hamiltonian. The block-tridiagonal matrix $\left( E + \mathrm{i} 0^+
-\mathcal{H}_0 - \Sigma \left( E \right) \right)$ can therefore be inverted
numerically using a highly convergent renormalization-decimation
algorithm\ \cite{sancho-hcsftcobasgf1985,nemec-qticn2007}, allowing us to go
beyond the energy range near the Fermi energy, where the special band
structure allows analytic inversion.

Starting with $\Sigma = 0$, each numerical iteration of this recursive
relation is equivalent to one additional perturbative order. Typically,
convergence is achieved after less then ten iterations, except for energies
near a van Hove singularity, where hundreds of iterations may be necessary.
This clearly indicates that low-order perturbation theory breaks down near
band edges.

The {\abbr{LDOS}} of each orbital $i$ in the unit cell can now be obtained
directly from the imaginary part of the Green function
\begin{eqnarray}
  g_i \left( E \right) & = & - \frac{1}{\pi} \mathrm{Im} \left[ \left( E +
  \mathrm{i} 0^+ -\mathcal{H}_0 - \Sigma \left( E \right) \right)^{- 1}
  \right]_{i, i} .  \label{g(E)}
\end{eqnarray}
Figs. \ref{fig:CNT-numerically} and \ref{fig:GNR-numerically} show this
quantity in direct comparison with the numerically exact value obtained by
sample averaging. The slight deviation visible at the flanks of the van~Hove
singularites is caused by the {\abbr{NCA}}\ \cite{hgle-vhsidmqwan2002}. The
elastic mean free path $\ell_{\mathrm{el}}$ based on the DOS of a disordered
system, as it is displayed in Fig.~\ref{fig:GNR-numerically} is no longer a
purely perturbative quantity, but it takes into account the scattering into
localized states present at any given energy.


The number of channels $N_{\mathrm{ch}}$ in Eqs.~(\ref{l_loc-Beenakker}) and
(\ref{l_el}) is another quantity that has to be reconsidered in the vicinity
of band edges. In a disorder-free quantum wire, $N_{\mathrm{ch}}$ is an integer
valued step function that has discontinuities at band edges. The numerically
obtained localization length, on the other hand, does not display any
discontinuities, so the discontinuities of $N_{\mathrm{ch}}$ have to be somehow
smoothed out by the disorder.

In any periodic system, the number of channels is equivalent to the
transmission through a cross section. For a Hermitian Hamiltonian, this
quantity will always take on integer values. Adding a complex self energy may,
however, result in non-integer values and smooth out the discontinuities of
the transmission function. Using the self-energy obtained from
Eq.~(\ref{self-energy}) allows thus to define an equivalent of the number of
channels $N_{\mathrm{ch}}$ for an Anderson disordered quantum wire.

The definition of the transmission through a cross section of a quantum wire
in the presence of a complex self-energy requires some care. Unlike a
Hermitian periodic wire, where the transmission through a cross section is
equal to that through a finite-length section, the imaginary part of the
self-energy would act as dissipative term in a conducting region and make the
result dependent on its length. To actually obtain the transmission through a
cross section, we should rather split the periodic quantum wire in only two
parts -- similar to the setup of a tunneling junction -- writing
\begin{eqnarray}
  \mathcal{H} & = & \left( \begin{array}{cc}
    \mathcal{H}_{\mathrm{L}} & 0\\
    0 & \mathcal{H}_{\mathrm{R}}
  \end{array} \right) + \left( \begin{array}{cc}
    0 & \mathcal{T}\\
    \mathcal{T}^{\dag} & 0
  \end{array} \right) =\mathcal{H}_0 +\mathcal{V} 
\end{eqnarray}
with two semi-infinite parts $\mathcal{H}_{\mathrm{L}}$ and
$\mathcal{H}_{\mathrm{R}}$ and the hopping matrix $\mathcal{T}$ that connects
both parts. Let us denote the Green functions of the two isolated
semi-infinite leads by $\mathcal{G}_{\mathrm{L}}$ and
$\mathcal{G}_{\mathrm{R}}$. The corresponding spectral functions are
$\mathcal{A}_{\mathrm{L / R}} = \mathrm{i} \mathcal{G}_{\mathrm{L /
R}}^{\mathrm{r}} - \mathrm{i} \mathcal{G}_{\mathrm{L / R}}^{\mathrm{a}}$ and the
Green functions of the complete system $\mathcal{G}^{\mathrm{r / a}} = \left(
E \pm \mathrm{i} 0^+ -\mathcal{H} \right)^{- 1}$. If $\mathcal{V}$ were a small
perturbation like in a tunneling junction its matrix elements $\left\langle l
| \mathcal{V} | r \right\rangle$ would directly give the transition amplitude
from an eigenmode $\mathcal{H}_{\mathrm{R}} \left| r \right\rangle = E \left|
r \right\rangle$ on the righthand side to one $\mathcal{H}_{\mathrm{L}} \left|
l \right\rangle = E \left| l \right\rangle$ on the lefthand side. In a
periodic system, however, $\mathcal{V}$ is of the same magnitude as
$\mathcal{H}_0$, so we need to consider the full expansion $\left\langle l |
\mathcal{V}+\mathcal{V}\mathcal{G}^{\mathrm{r}} \mathcal{V} | r
\right\rangle$, generalizing the tunneling-transmission $\mathrm{tr} \left(
\mathcal{A}_{\mathrm{L}} \mathcal{V}\mathcal{A}_{\mathrm{R}} \mathcal{V}
\right)$ to the nonperturbative expression\ \cite{nemec-qticn2007}:
\begin{eqnarray}
  T & = & \mathrm{tr} \left[ \mathcal{A}_{\mathrm{L}} \left(
  \mathcal{V}+\mathcal{V}\mathcal{G}^{\mathrm{r}} \mathcal{V} \right)
  \mathcal{A}_{\mathrm{R}} \left(
  \mathcal{V}+\mathcal{V}\mathcal{G}^{\mathrm{a}} \mathcal{V} \right) \right]
  .  \label{transmission}
\end{eqnarray}
Note that in the case of a finite self-energy $\Sigma$ all along the periodic
wire, this expression differs substantially from more commonly used approaches
that depend on the definition of a finite-length conductor between two leads.
Numerically, we can exploit the finite support of $\mathcal{V}$ by computing
only the contacted regions of $\mathcal{G}_{\mathrm{L}}$ and
$\mathcal{G}_{\mathrm{R}}$\ \cite{sancho-hcsftcobasgf1985,nemec-qticn2007}.



The localization length $\ell_{\mathrm{loc}}$ of armchair {\abbr{CNT}}s at
$E_{\mathrm{F}}$ can be derived analytically directly from $\ell_{\mathrm{el}}$
and $N_{\mathrm{ch}}$\ \cite{jiang-uefllimcn2001}. For general energies,
$\ell_{\mathrm{loc}}$ has so far only been accessible by numerical
means\ \cite{avriller-cdsicnmtoqtr2006,biel-alicnddate2005}. Based on our
expressions for the {\abbr{DOS}} and the number of channels in a
homogeneously disordered quantum wire, it is now possible to obtain an
approximate value for $\ell_{\mathrm{loc}}$ in the whole energy range. As can be
seen in Fig.~\ref{fig:CNT-numerically}, the analytical value for
$\ell_{\mathrm{loc}}$ agrees with the numerically exact results fairly well. The
remaining deviation can be understood in view of the oversimplified nature of
Eq.~(\ref{l_loc-Beenakker}), which is based on a model with a fixed number of
equivalent channels with a single value $\ell_{\mathrm{el}}$ for all scattering
processes.

\begin{figure*}
  \resizebox{1.0\columnwidth}{!}{\includegraphics{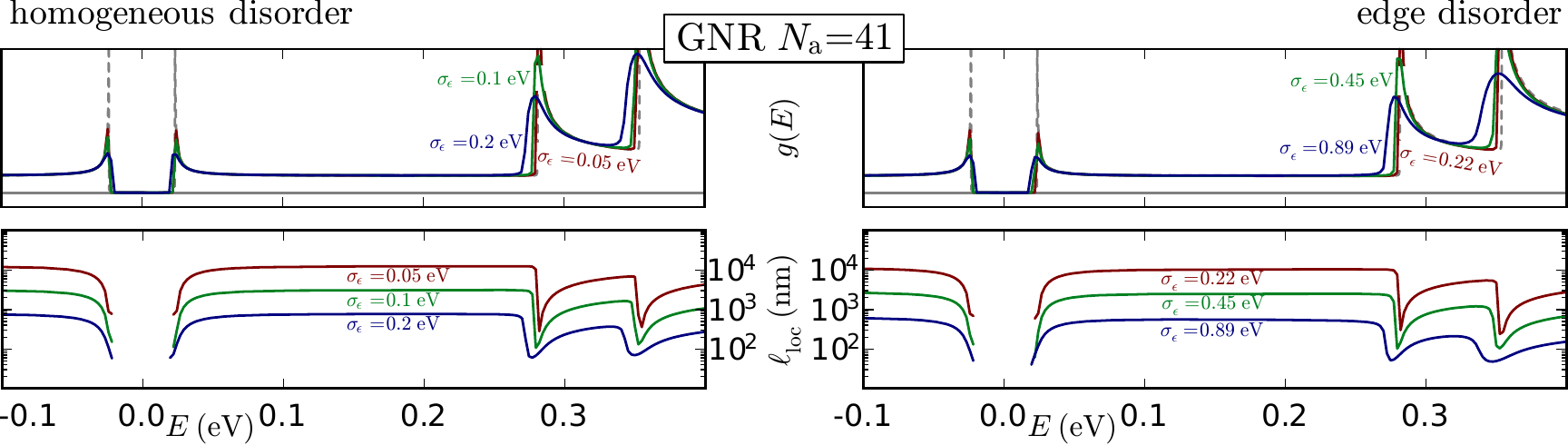}}
  \caption{\label{fig:GNR-analytically}Analytically obtained quantities for an
  armchair edge GNR of width $N_z = 41$ with Anderson disorder of various
  strengths. Dashed lines: Density of states $g \left( E \right)$ of the clean
  system. Solid lines: $g \left( E \right)$ and localization length
  $\ell_{\mathrm{loc}}$ obtained analytically. Left: homogeneous disorder over
  all atoms across the {\abbr{GNR}}. Right: disorder concentrated on the
  outermost atoms along both edges. The disorder strength
  $\sigma_{\varepsilon_i}$ per atom is scaled such that the square total
  $\Sigma_i \sigma_{\varepsilon_i}^2$ is the same for both types of disorder.}
\end{figure*}

Armchair edge {\abbr{GNR}}s can be understood as unrolled zigzag
{\abbr{CNT}}s and are indeed physically very comparable. Similar to
semi-metallic $(3 N, 0)$ nanotubes, which are metallic in zone-folding
approximation and develop a small gap due to their curvature, ribbons of width
$N_a = 3 M - 1$ (counted in rows of carbon dimers) have a tiny gap that is
caused by edge effects only. In Fig.~\ref{fig:GNR-numerically} one can see
that the {\abbr{DOS}} is smoothed out and the van Hove singularities shifted
in the same way as it was observed in {\abbr{CNT}}s. Near $E_{\mathrm{F}}$,
this leads to a narrowing of the gap which is well reproduced by the data
based on Eq.~(\ref{g(E)}). The states near the gap exhibit a short
localization length, as it is expected from the fact that they are caused by
the disorder within the forbidden energy range of the clean system. The
analytically obtained localization length coincides with the numerical value
to a procesion comparable to the case of {\abbr{CNT}}s. Again, the true
value is slightly underestimated when several inequivalent channels are
present at the same energy.

Reducing the strength of the disorder to realistic values leads to a
significant reduction of the smoothing effect on van Hove singularieties.
Besides this effect, the localization lengths simply scale with $1 /
\sigma_{\varepsilon}^2$ wherever band edges have no influence, as can be seen
in Fig.~\ref{fig:GNR-analytically}. Assuming the amount of intrinsic disorder
in {\abbr{GNR}}s to be similar to that measured in {\abbr{CNT}}s at
$\sigma_{\varepsilon} \sim 0.05 \hspace{0.2em} \mathrm{eV}$, we find a
localization length of $\ell_{\mathrm{loc}} \sim 13 \hspace{0.2em} \mu \mathrm{m}$
in armchair {\abbr{GNR}}s of $5 \hspace{0.2em} \mathrm{nm}$ width within the
plateau of only two transmitting channels at sufficient distance to the gap.

The edges of a {\abbr{GNR}} must be considered to be especially sensitive to
disorder due to the lower mechanical rigidity and higher chemical activity. We
therefore studied the effect of pure edge disorder as depicted in the right
panels of Fig.~\ref{fig:GNR-analytically}. As can be immediately seen, the
overall effect of the disorder scales with the square total $\Sigma_i
\sigma_{\varepsilon_i}^2$ over the disorder strength of all individual atoms.
When scaling the disorder strength in this way, the effects of edge disorder
are identical to the case of homogeneous disorder except for the band edges,
where the van Hove singularities are smeared out stronger in the case of pure
edge disorder.

In zigzag edge {\abbr{GNR}}s the edge states at the Fermi energy add a
considerable complication. Numerical studies have been done before on these
systems using both Anderson-type disorder of short or long
range\ \cite{wakabayashi-pccaucidgn2007,lherbier-tlsidgmslrade2008} or edge
defects\ \cite{li-qcognwed2008,gunlycke-sgnwed2007,wimmer-stirgn2008}. We
found that the {\abbr{DOS}} obtained by Eq.~(\ref{g(E)}) shows a
considerable deviation from the true value obtained numerically. Possibly,
this effect is related to the highly anomalous behavior found in
two-dimensional sheets of graphene\ \cite{peres-epotc2006}. Trying to obtain
$\ell_{\mathrm{loc}}$ via Eq.~(\ref{l_loc-Beenakker}) is bound to fail even more
seriously as this approach completely neglects the different mean free paths
and the spatial separation between the edge channels and the bulk channel. In
fact, though this approach correctly describes a very short localization
length around the edge states, it failes to reproduce the correct scaling with
the ribbon width.


To conclude, we have demonstrated that our generalized expression for the
density of states in Anderson-disordered quantum wires in combination with a
highly convergent numerical evaluation scheme does reproduce to good precision
the true data obtained by performing a numerical sample-average. The method
allows to efficiently explore of the full energy range and can be applied to
arbitrary quantum wires, including {\abbr{CNT}}s and armchair-edge
{\abbr{GNR}}s. We further demonstrated how to obtain the localization length
of those systems from the density of states in the full energy range,
including the vicinity of van Hove singularities which are notoriously hard to
access by analytical means. Again, the resulting data shows good agreement
with statistically obtained values. The case of zigzag-edge {\abbr{GNR}}s
states has to be excluded due to the extreme inequivalence of the different
channels close to the Fermi energy and remains a problem for the future.


We acknowledge fruitful discussions with I.~Adagideli, R.~Guti\'errez,
R.~R\"omer and M.~Wimmer as well as the cooperation with T.~Breu and
V.~Agarval at an early stage of the project. This work was funded by the
Volkswagen Foundation under Grant No.~I/78~340, by the Deutsche
Forschungsgesellschaft (GRK~638) and by the European Union program CARDEQ
under Contract No.~IST-021285-2. Support from the Vielberth Foundation is also
gratefully acknowledged.

\section*{References}

\bibliographystyle{unsrt}

\end{document}